# Tracing dynamics of laser-induced fields on ultra-thin foils using complementary imaging with streak deflectometry

F. Abicht[1,*], J. Bränzel[1], G. Priebe[2], Ch. Koschitzki[1], A. A. Andreev[1,3,4], P.V. Nickles[1,5], W. Sander[1,6,7,†] and M. Schnürer[1,‡]

[1]*Max-Born-Institut, Max-Born-Str. 2a, 12489 Berlin, Germany*

[2]*XFEL GmbH, Notkestr. 85, 22607 Hamburg, Germany*

[3]*Vavilov State Optical Institut, Birzhevaya line 12, 199064 St. Petersburg, Russia*

[4]*St. Petersburg University, 199064 St. Petersburg, University emb.6, Russia*

[5]*Center of Relativistic Laser Science, Institute for Basic Science, Gwangju 500-712. Rep. of Korea*

[6]*Technical University Berlin, Straße des 17. Juni 135, 10623 Berlin, Germany*

[7]*ELI-DC International Association AISBL*

We present a detailed study of the electric and magnetic fields, which are created on plasma vacuum interfaces as a result of highly intense laser-matter-interactions. For the field generation ultra-thin polymer foils ($30 - 50$ nm) were irradiated with high intensity femtosecond ($10^{19} - 10^{20}$ W/cm$^2$) and picosecond ($\sim 10^{17}$ W/cm$^2$) laser pulses with ultra-high contrast ($10^{10} - 10^{11}$). To determine the temporal evolution and the spatial distribution of these fields the proton streak deflectometry method has been developed further and applied in two different imaging configurations. It enabled us to gather complementary information about the investigated field structure, in particular about the influence of different field components (parallel and normal to the target surface) and the impact of a moving ion front. The applied ultra-high laser contrast significantly increased the reproducibility of the experiment and improved the accuracy of the imaging method. In order to explain the experimental observations, which were obtained by applying ultra-short laser pulses, two different analytical models have been studied in detail. Their ability to reproduce the streak deflectometry measurements was tested on the basis of three-dimensional particle simulations. A modification and combination of the two models allowed for an extensive and accurate reproduction of the experimental results in both imaging configurations. The controlled change of the laser pulse duration from 50 femtoseconds to 2.7 picoseconds led to a transition of the dominating force acting on the probing proton beam at the rear side of the polymer foil. In the picosecond case the ($\vec{v} \, x \, \vec{B}$)-term of the Lorentz force dominated over the counteracting $\vec{E}$-field and was responsible for the direction of the net force. The applied proton deflectometry method allowed for an unambiguous determination of the magnetic field polarity at the rear side of the ultra-thin foil.

---

[*] Corresponding author, electronic mail: abicht@mbi-berlin.de
[†] In memoriam Wolfgang Sandner, deceased December 5, 2015
[‡] Corresponding author, electronic mail: schnuerer@mbi-berlin.de

# I. INTRODUCTION

Laser driven particle acceleration in plasmas is a well-established and still growing field of research. The possibility to accelerate particles over short distances allows for the construction of compact accelerators that have significant benefits for a variety of applications [1], as for example in basic sciences, material research and in the bio-medical sector. The lifetime and dynamics of the acceleration field depends on energy dissipation processes and thus is coupled to the duration of the laser pulse. Typically the time scale of these processes extends over several times the pulse duration, which results in highly transient acceleration field structures. Characterization of these fields is not only important to conclude about the plasma kinematics involved in the acceleration process but also to investigate the potential of cascaded acceleration schemes and the application of charged particle injection.

Here we study the strong fields on plasma vacuum interfaces generated on thin foils when irradiated with ultra-high intensity femtosecond ($10^{19} - 10^{20}$ W/cm$^2$) and picosecond ($\sim 10^{17}$ W/cm$^2$) laser pulses.

The focus of this investigation is not only on the target normal component of the electric field, which plays the key role in Target Normal Sheath Acceleration (TNSA) [2], but also on the radial electric field component (parallel to the foil surface) and the occurring toroidal magnetic field.

The present study develops new aspects to gain a comprehensive view on strong accelerating and deflecting fields in laser ion acceleration. A favorable combination of laser and target parameters was applied: for the generation of the investigated field structures ultra-thin polymer foils ($30 - 50$ nm) [3,4] and ultra-high contrast ($10^{10} - 10^{11}$) laser pulses were used. As previous experiments [5] have shown, these conditions allow for an optimization of the TNSA-process [6,7]. In addition thin foils reduce the scattering of penetrating probe particles, which is a prerequisite for the applied proton imaging method [8].

To investigate the temporal evolution and the spatial distribution of the fields the proton streak deflectometry method [5] has been used and developed further. By applying the method in two imaging configurations we became access to complementary information about the investigated field structure.

In this context the high laser contrast had a substantial influence: it increased the quality of the probing proton beam and led to a high reproducibility of the investigated field structure and hence of the whole experiment. In comparison with former experiments [5], consecutive proton imaging measurements now exhibited relatively low fluctuations when experimental conditions were kept constant. Therefore, the combination and comparison of images with different experimental parameters enables a qualitative and quantitative description of how the change of a specific parameter impacts on the laser induced field distribution.

In this paper several different experimental parameters were changed in consecutive measurements in order to monitor the associated induced change. The foil position with regard to the probing proton beam was varied systematically. This way the spatial extension of the fields could be investigated without decreasing the temporal and spatial resolution. A change of the laser pulse duration from the femtosecond to the picosecond range led to a transition of the dominating field action on the probing proton beam. In contrast to the femtosecond domain, in the picosecond domain the influence of the magnetic field was counter-acting to the direction of the electric field and even more effective. Thus, the applied proton deflectometry method allowed for an unambiguous determination of the magnetic field polarity, which can provide insight to the field generating mechanisms. To gain such information using two-dimensional-imaging with different pump-probe-delays needs much more effort.

In order to explain the experiments with femtosecond laser pulses, two different analytical models have been studied in detail. Their ability to reproduce the streak deflectometry measurements was tested on the basis of three-dimensional particle simulations by including the corresponding analytic field descriptions and accounting for the specific experimental conditions. As a result, specific but different characteristic features of the streak deflectometry measurements could be explained. Due to complementary streak deflectometry measurements with two different imaging configurations and using different experimental parameters a comprehensive investigation of the field structure becomes possible, such as the signature of a propagating proton front and the influence of the different magnetic and electric field components. This was not only useful to determine the application range and limitations of the single models, but also for the further development of the model descriptions itself. A modification and combination of the two models allowed for an extensive and accurate reproduction of the experimental results. This way, not only single and selected measurements could be reproduced by the particle simulation, but also (and at the same time) measurements with varied experimental parameters and measurements that were obtained using different imaging configurations. In this regard, the described methodological approach offers a new path for a comprehensive reconstruction of the spatial and temporal field distribution of laser-plasma interactions.

## II. EXPERIMENTAL SETUP

The experiments have been conducted with the "High Field Laser" system at Max-Born-Institute Berlin. It consists of two separate but optically synchronized Ti:Sapph amplifier chains, which are seeded with a shared XPW-frontend [9]. Due to the different architecture of these amplifier chains (arm A and B) laser pulses with different pulse duration and temporal contrast are generated. Laser arm A (100 TW) is based on a regenerative amplifier (inclusive spectral steering) and delivers pulses with 25 fs (FWHM of intensity) duration at an ASE background level of $10^{-10} - 10^{-9}$. Laser arm B (70 TW) consists only of multi-pass amplifier sections and delivers pulses with 35 fs duration at an ASE background level of $10^{-11} - 10^{-10}$.

In the experiments described here, the (pump) laser pulses of laser arm B are used to generate a laser-matter interaction on a 30 nm polymer ($C_5H_7O_2$) foil [3,4] with a solid state electron density $n_{e0}$ of $3 \times 10^{23}$ $1/cm^3$. The associated and fast evolving field structure of this interaction is investigated by means of a (probing) proton beam that is created with the help of laser arm A. This "probing" proton beam stems from laser irradiated titanium foils (5 μm thick) and its adherent CH-contamination.

In the presented cases, the focal intensity of the probe laser pulse was $\sim 10^{19}$ W/cm$^2$. The focal intensity of the pump laser pulses depended on the setting of the laser pulse duration, which was varied between consecutive measurements by changing the position of the grating compressor. Here, two cases are discussed. In the "femtosecond case" a 50 fs laser pulse with an intensity of $\sim 2 \times 10^{19}$ W/cm$^2$ was applied and in the "picosecond case" a pulse length of 2.7 ps was used leading to an intensity level of $\sim 6 \times 10^{17}$ W/cm$^2$. The intensity calculations are based on measurements of pulse energies, pulse durations and focal spot-sizes with encircled energy content. The pulse energies are measured in front of the grating compressors. On target values are calculated with regard to the measured relative transmissions of the grating compressors and beam-lines inclusive focusing optics.

The experimental setups in transversal and longitudinal imaging configuration are illustrated in Fig. 1(a) and Fig. 1(b), respectively. The only difference between both setups consists in a rotation of the interaction (polymer) foil by 90° while the incident angle of the pump laser pulse remains constant. Thus the angle of incidence is changed from 45° to −45° in reference to the target normal. Therefore the orientation of the laser induced field distribution in space is both rotated (by 90° around the y-axis) and mirrored (along the x-axis). However, if the laser parameters are identical in both cases, the induced field structure itself is not changed, only its absolute orientation in space. On their way to the detector the probing protons propagate either parallel (longitudinal configuration) or perpendicular (transversal configuration) to the surface normal of the polymer foil (interaction target) and are deflected by the induced fields due to the acting Lorentz force. By inserting of a beam mask (grating) the deflection along the x-direction becomes detectable. However, depending on the imaging configuration the induced proton deflection is dominated by different components, as illustrated in Fig. 1(a) and Fig. 1(b). Here $E_N$ indicates the main electric field component normal to the target surface, $E_R$ the radial electric field component and $B$ the toroidal magnetic field. By measuring the energy dependent proton deflection in both imaging configurations complementary information on the fast evolving field structure is obtained. For this purpose a modified Thomson spectrometer [5] was applied. This method is called "proton streak deflectometry" [5] and is explained in detail in Refs. [5,10].

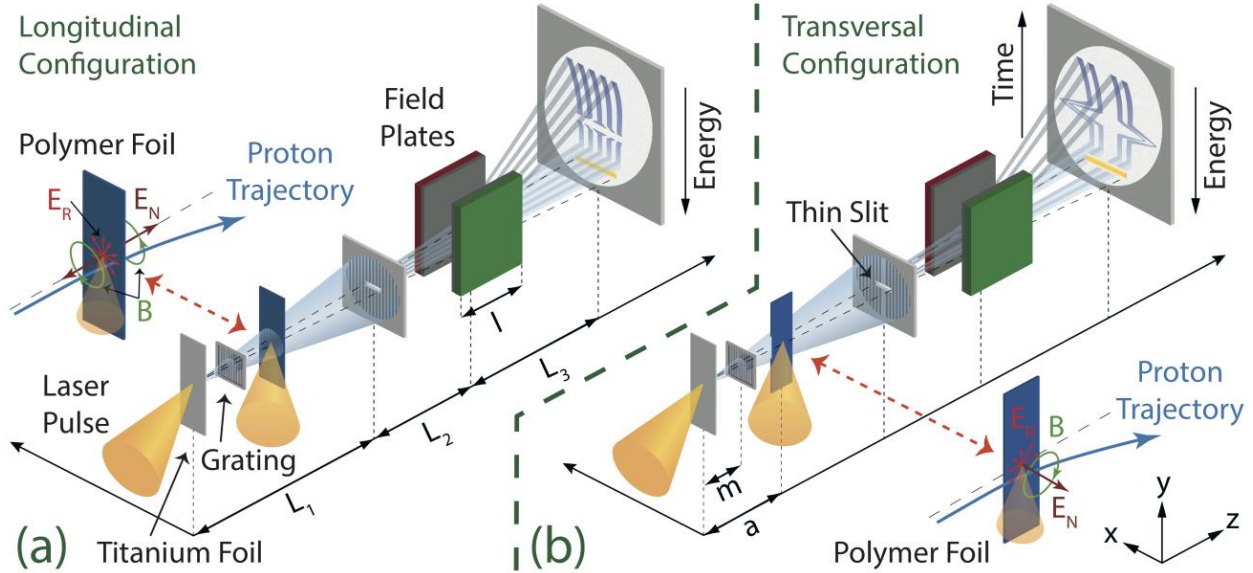

FIG. 1. Setup and imaging configurations: (a) longitudinal configuration, (b) transversal configuration.

The entrance slit has a width of 288 μm and a length of about 1 cm. The used distances in Fig. 1 are: $L_1 = 477$ mm, $L_2 = 228$ mm, $L_3 = 515$ mm, $a = 35$ mm, $m = 20$ mm. The applied magnet has the length $l = 50$ mm and corresponding to this extension the effective field strength $B_x = 0.34$ T. As a detector a multi-channel plate (MCP) from Hamamatsu (Type F1942-04) coupled with a phosphor screen was used in combination with an attached camera. The imaging quality of the MCP was verified with laser accelerated proton beams and projection imaging of test objects and the energy dependent ion sensitivity was calibrated [11]. The overall energy resolution $\Delta\epsilon$ of the imaging system was approximately 45 keV for proton energies around 2.5 MeV. In general the relative energy resolution is better than 3% of the relevant kinetic energy values [10]. In order to exclude contributions of other ion species besides protons the MCP was temporally gated. By switching off the voltage between the phosphor screen and the MCP, at a time 130 ns after the laser-plasma interaction, a pure proton signal with energies above 0.5 MeV could be detected.

In transversal configuration, the temporal resolution $\Delta t$ is around 30 ps if an interaction length of $l_l$ of 100 μm is assumed [10]. In this configuration two additional Thomson spectrometers were installed opposite the front and rear surface of the polymer foil. These spectrometers were applied to record the energy distribution of the individual ion species that are accelerated in opposite directions normal to the surface of the interaction target.

### III. EXPERIMENTAL RESULTS

Figs. 2(a) - 2(c) show a selection of processed streak deflectometry measurements, which were obtained in transversal imaging configuration. Different regions of the laser-induced fields were probed by changing the x-position of the polymer foil with regard to the probing proton beam in consecutive shots. The measured proton density distributions $\rho(x, y)$ on the MCP have been processed by means of a numerical coordinate transformation method that is presented in detail in Ref. [10] As a result the distributions $\tilde{\rho}(\beta, t_p)$ can be visualized as a function of the initial proton emission angle $\beta$ (in the x-z-plane) and the probing time $t_p$, which are represented by the upper abscissa and the left ordinate, respectively. The lower abscissa indicates the corresponding proton energy for $\beta = 0$. The right ordinate shows the relative lateral distance $x_f$ (in x-direction) between a proton and the focus position at the moment in time, when the proton passes trough the x-y-plane at z=a [cf. Fig. 1(a)]. The lateral focus distance $x_f$

depends both on the initial emission angle of the proton and the x-position of the laser focus on the polymer foil. However, the definition of the lateral focus distance $x_f$ holds only for an idealized proton, which is not deflected by the laser-induced fields. Due to the action of the Lorentz force, the trajectory of a proton can be altered in the extended field region around the polymer foil. Thus the final propagation angle of a proton (in the x-z-plane) might differ from its initial ejection angle $\beta$. For this reason the lateral focus distance $x_f$ is only an approximation of the real distance, if a proton is deflected in the laser induced field region around z=a.

The arrival of the laser pulse at the titanium target initiates the creation of the probing proton beam and defines the absolute time zero. By means of an optical delay stage within the experimental chamber the arrival time $t_{pump}$ of the laser pump pulse at the polymer target was adjustable with respect to time zero [10]. Thus, the time $t_{pump}$ defines the arrival time of the pump laser pulse at the polymer target with regard to the creation of the probing proton beam at the titanium target. In addition, the time $t_{pump}$ can be regarded as the arrival time of protons at the polymer target, which are accelerated at the titanium target at time zero and propagate with the specific energy $\epsilon = \epsilon_{pump}$. In fact this relation is the definition of the so-called t-pump energy $\epsilon_{pump}$. In other words a proton with the kinetic energy $\epsilon = \epsilon_{pump}$ needs the time $t_p = t_{pump}$ to propagate from the titanium to the polymer target.

In the presented experiments the time $t_{pump}$ is set to 1571 ps, which corresponds to the proton t-pump energy $\epsilon_{pump} = 2.6$ MeV. With the generation of the electric fields, the probing protons (with energies around $\epsilon \approx \epsilon_{pump}$) are deflected away from the surfaces of the interaction target depending on their relative position, propagation direction and energy. The temporal dependence of the deflection can be divided in several intervals.

Protons with energies $\epsilon > \epsilon_{pump}$ reach the polymer target ($z = a$) even before the laser pump pulse impinges on its surface. The initiated fields only affect a proton if its relative distance to the interaction center (pump focus) is less than the radial extension of the field distribution in z-direction.

In order to visualize the influence of transversal force components (in x-direction, i.e. perpendicular to the propagation direction of the probing proton beam) a grating with 300 LPI (lines per inch) was used to intersect the proton beam (cf. Figure 1) into single beamlets. Each of these beamlets consists of protons that are ordered in time and space (in propagation direction) according to their energy $\epsilon$. In addition, each beamlet can be attributed to protons having initial ejection angles of a certain interval. Depending on their energy, the final propagation angles of these protons can be changed during the interaction with the laser-induced fields. Correspondingly, the mean propagation angle of a whole beamlet can be altered for a certain energy interval and might differ from its initial ejection angle $\beta$. The thin slit confines the beamlets further, allowing only for protons propagating close to the x-z-plane. Thus, only deflections in x-direction can be measured. Within the permanent magnetic field of the Thomson spectrometer protons are dispersed in y-direction according to their energy. On the detector plane (x-y-plane), the energy distribution of each beamlet is transformed into a spatial distribution along y-direction and becomes visible in form of proton traces (proton streaks). The time dependent action of transverse force components leads to deflections of these traces, i.e. streak deflections. Regarding the processed images of Figs. 2, 4 and 5, the streak deflections are shown as deviations from the initial emission angle $\beta$. The value of the initial emission angle $\beta$ that corresponds to a certain proton trace is given by its relatively constant position for energies $\epsilon \gg \epsilon_{pump}$. These energies correspond to probing times ($t_p \ll t_{pump}$) well before the deflecting fields are initiated. Thus, the propagation angles of protons with such energies are not altered by these fields and are equal to their initial emission angles. The dependence of the initial emission angle on energy during the acceleration process at the titanium foil has been investigated in Ref. [5]. For the energy range considered here, this dependency in negligible with regard to the induced deflections around the polymer foil.

In view of all measurements, the deflected traces exhibit a pronounced deflection symmetry, indicating a highly symmetric field distribution. Protons passing the interaction target on different sides [cf. Figs. 2(a) and 2(b)], but with similar energies $\epsilon_p$ and similar initial absolute values of the lateral focus distance $|x_F|$, show similar deflection amplitudes. On both sides the deflection strongly depends on the initial distance to the target surface. Protons passing closer to the foil surface are deflected stronger than protons, which pass at a larger distance. The time $t_{pump}$ defines the arrival time of the pump laser pulse at the polymer foil with regard to the creation of the probing proton beam at the titanium target.

As shown in Figs. 2(c) and 2(d) the deflection becomes visible at the time $t_f = t_p = 1410$ ps and can be characterized by a steep increase. Proton traces with lateral focus distances $|x_F| \lesssim 650$ μm (for $t_p \leq t_{pump}$) show a deflection maximum at the time moment, which corresponds exactly to the setting of the pump time $t_{pump} = 1571$ ps. For traces at higher distances ($|x_F| \gtrsim 650$ μm) the deflection maximum is shifted towards lower proton energies. After the maximum the deflection decays relatively quickly for traces with $|x_F| \lesssim 650$ μm, on a time scale of approximately 80 ps. Then the decay flattens and within 250 ps the deflection declines to a quasi-constant value. At higher distances ($|x_F| \gtrsim 400$ μm) the deflected traces exhibit a second local maximum before dropping to constant values in time. At this point the proton deflection becomes relatively independent on the initial ejection angle $\beta$. As explained in Ref. [10] the time $t_f \approx 1410$ ps when the deflection becomes visible is connected with the energy $\epsilon_f \approx 3.23$ MeV of the fastest influenced protons, as illustrated in Fig. 2(c).

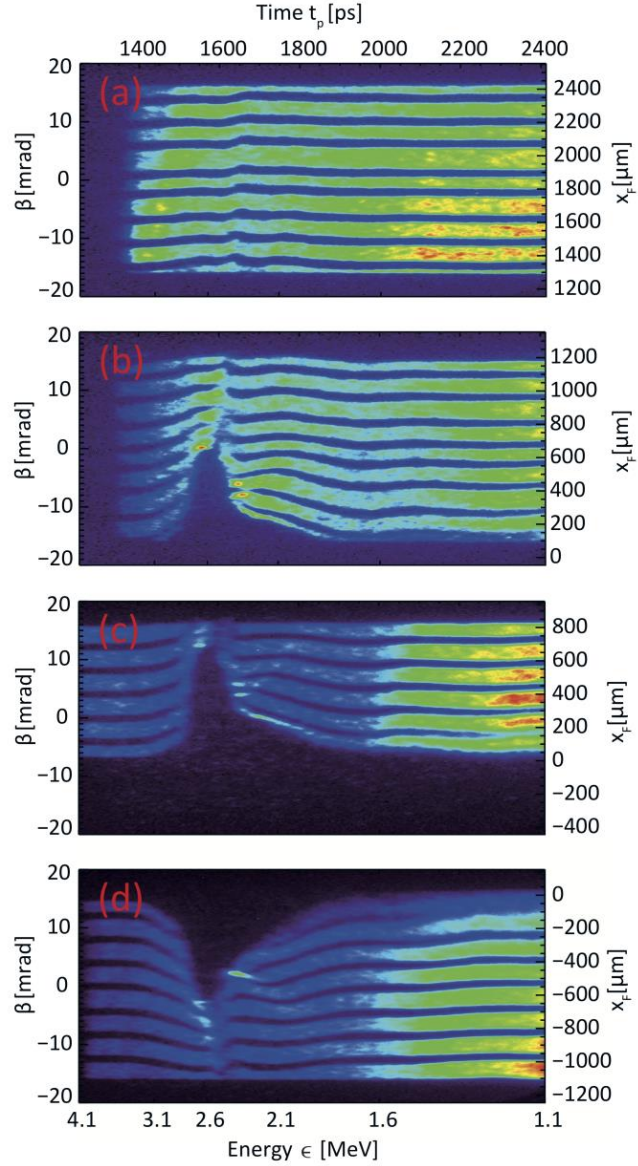

FIG. 2. Proton streak deflectometry measurements in transversal imaging configuration. The x-position ($d_x$) of the polymer foil with regard to the probing proton beam was changed in consecutive shots: (a) $d_x = -1854$ μm, (b) $d_x = -654$ μm, (c) $d_x = -254$ μm, (d) $d_x = 545$ μm.

The corresponding velocity value can be used to estimate the extension of the electric field in z-direction under the assumption that the electric field component in z-direction (along the surface) is negligible. In the case of Fig. 2(c) an extension radius $r_f$ of 4.3 mm is obtained. Considering that the foil target has a total extension of 5 mm in z-direction and is irradiated in its center, the deduced extension radius is relatively high. This indicates that radial field components along the foil surface (in z-direction) cannot be neglected in the presented case. Using the extension of the foil in z-direction as an upper limit of the radial field extension, the time resolution of the method can be estimated [10] At the probing time $t_p = t_{pump}$ the time resolution $\Delta t$ is approximately 250 ps. Simultaneously registered ion spectra from front and rear side emission of the polymer foil target are nearly identical [10] and confirm the symmetry of Fig. 2(c) and 2(d).

Figs. 2(a) and 2(b) show the recorded streak deflections, which correspond to protons passing the foil surface at larger distances at the time $t_{pump}$, i.e. for higher values of $|x_F|$. Proton streak deflections are visible up to the maximum-recorded value $|x_F| \approx 2400$ µm, which is a measure of the effective field extension normal to the foil surface. Thus, for the electric field component $E_N$ normal to the target surface an effective extension length $l_{En} \gtrsim 2400$ µm is found. For distances $|x_F| \gtrsim 700$ µm the recorded proton streaks do not overlay and are clearly distinguishable, even at the deflection maximum. For these distances the deflection amplitude is smaller than the width of a proton streak ($\approx 50$ µm) and decreases at higher values of $|x_F|$. This dependency could be explained by a field gradient $\partial E/\partial x$, which becomes relatively flat for larger distances ($|x_F| \gtrsim 700$ µm). If protons of adjacent streaks are affected by almost the same force, the amplitude of their deflection becomes similar and different streaks cannot overlay on the detector. A careful analysis [10] of the presented deflection measurements shows that the deflection maximum is shifted to higher probing times $t_p$ with increasing distance $|x_F|$. Assuming the field of an expanding ion front causes the deflection maximum, the slope of this shift can be regarded as the expansion velocity of the front. In case of Fig. 2(a) a velocity of $2.39 \times 10^7$ m/s is deduced, which corresponds to a proton energy of 3.01 MeV. In case of Fig. 4(c) the velocity $2.34 \times 10^7$ m/s is obtained, which results in a proton energy of 2.87 MeV. A comparison with the associated proton energy spectra shows that these values are in good agreement with the measured cut-off energies [10], which confirms the assumption.

Fig. 5(b) shows a selected streak deflectometry measurement in longitudinal imaging configuration. The time $t_{pump}$ was set in agreement with the proton t-pump energy $\epsilon_{pump} = 2.4$ MeV.

As before, the measured proton density distribution $\rho(x, y)$ on the detector was processed using the numerical coordinate transformation [10]. However, in this case the proton density distributions $\tilde{\rho}(\beta, \epsilon)$ is not shown depending on the probing time $t_p$, but as a function of the proton energy $\epsilon$ (lower abscissa). Note, that in Fig. 5 the proton energy $\epsilon$ increases in reverse direction as opposed to Fig. 2. The initial emission angle $\beta$ and the lateral focus distance $x_F$ are indicated on the left and right ordinate, respectively.

In order to visualize the influence of transversal force components (along the foil surface) also a grating with 200 LPI was used to intersect the probing proton beam. The resulting proton streaks [Fig. 5(b)] on the detector exhibit prominent density dips in the energy range of 2.4 MeV, which corresponds to the selected proton t-pump energy $\epsilon_{pump}$. These dips can be explained as a consequence of the de- and accelerating action of the normal electric field component $E_N$. Thus the action of the ambipolar electric field leads to a redistribution of the energies within the probing proton beam [12]. In agreement with this interpretation, the dips are most pronounced near the center of the interaction ($|x_F| = 0$ µm).

In addition, a bending of the proton streaks is visible at both sides of the dips, indicating the presence of radial electric field components $E_R$ and/or the influence of a magnetic field $B$ [10]. On both sides the bending is directed outwards in respect to a line through the interaction center ($|x_F| = 0$ µm), however the deflection is stronger at the low energy side.

At greater distances to the interaction center the outward deflection of the protons decreases. As a result no bending of the streaks is detectable for $|x_F| \gtrsim 600$ µm at the high-energy side of the dips. This observation is supported by additional measurements where the position of the pump focus was displaced in x-direction [10]. These measurements show, that the bending of the streaks at the low energy side becomes smoother for greater values of $|x_F|$ and the deflection maximum is shifted towards lower energies [10]

# IV. MODEL APROACHES

In the following three theoretical models are proposed, which provide different analytical descriptions of laser-induced electric and magnetic fields on thin foils. For convenience these models are referred to as model A, model B and model C. In principle model A and model B are variants of already published model descriptions [13,14,15]. These descriptions have been extended and developed further in this work in order to explain the specific characteristic features of the streak deflectometry measurements in longitudinal and transversal imaging configuration. Model C is basically a combination of these two models and allows for an extensive and accurate reproduction of all experimental results.

## A. Model A

Model A is based on a published analytical model description [13]. It describes the electric and magnetic field distribution, which is generated on an ultra-thin foil (< 1 μm) if an ultra-short (< 50 fs) laser pulse irradiates the target surface at an angle of 45°. The model accounts for the generation of hot electrons and their motion through the target (along the target normal, z-axis) as well as their transverse expansion along the target surface (x-axis) in a two-dimensional Cartesian geometry.

The Field dynamics depends on the evolution of the positively charged region, which is induced on the target surface and governed by the charge conservation law. The expansion of hot electrons is balanced by a flow of cold electrons from the target periphery in direction of the charged region. As a result the normalized spot size $L_n(t)$ of the induced positive charge expands in transverse direction depending on the target conductivity $\sigma_{cold}$ and on the effective electron collision frequency $\nu_{ef}$. Its temporal evolution is described by the expression

$$L_n(t) = \sqrt{L_{n0}^2 + \frac{\kappa}{\nu_{ef}} \cdot \left(1 - \exp(-\nu_{ef}\, \omega_{pef}\, t)\right)}\,, \qquad (1)$$

where $L_{n0} = D_{e0}/r_{D0}$ and $\kappa = (\sigma_{cold}\, \Delta a)/(\omega_{pef}\, r_{D0})$. The initial spot size in real space is denoted with $D_{e0}$ and the thickness of the polymer foil with $\Delta a$. The (initial) plasma frequency of hot electrons is indicated by $\omega_{pef}$ and can be calculated depending on the initial density of hot electrons using the formula $\omega_{pef} = \sqrt{n_{e0} e^2/\varepsilon_0 m_e}$. Here $m_e$ is the mass and $e$ the charge of an electron. The initial Debye radius has the symbol $r_{D0}$ and depends both on the density $n_{e0}$ and the temperature $T_e$ of hot electrons using the equation $r_{D0} = \sqrt{\varepsilon_0 k_B T_e/e^2 n_{e0}}$.

The net positive charge density on the foil $\rho(x,z)$ is given by $\rho(x,z) = e n_{e0} \eta_{e0} \rho_\eta(x_n)\delta(z_n)$ where the coordinates $x_n = x/r_{D0}$ and $z_n = z/r_{D0}$ are normalized to the Debye radius of hot electrons $r_{D0}$. The sharp boundary of the surface charge in target normal direction is modeled using the Dirac-delta function $\delta(z_n)$. The transverse profile of the charge density is described by a one dimensional normalized distribution function $\rho_\eta(x_n)$ multiplied with the normalization factor $\eta_{e0} = 2\pi\, D_{e0}/r_{D0}$. In this paper a normalized Lorentzian distribution is applied

$$\rho(x_n, t) = \frac{1}{\pi S_L L_n(t)} \cdot \left[1 + \left(\frac{|x_n|}{S_L L_n(t)}\right)^2\right]^{-1}, \qquad (2)$$

where the constant factor $S_L$ ($\in [0,1]$) determines the (initial) width of the distribution in relation to $L_n(t)$.
Under the assumptions that the energy transfer between hot electrons and ions follows the adiabatic law and that the electron inertia can be neglected in the standard force equation for the density the two-dimensional Poisson equation can be solved [16]. Using the modified Bessel function of second kind $K_0$ the electric potential $\phi(x_n, z_n, t)$ can be evaluated. The resulting formula

$$\phi(x_n, z_n, t) = E_0 \frac{\eta_{e0}}{4\pi} \int \rho(\tilde{x}_n, L_n(t)) K_0 \left( \sqrt{\frac{z_n^2}{2} + \frac{x_n - \tilde{x}_n}{2}} \right) d\tilde{x}_n , \quad (3)$$

where $E_0 = \sqrt{2 I_L/\varepsilon_0 c}$, allows for the deduction of the electric and magnetic field distribution. However, these expressions also depend on modified Bessel functions, which make them unsuitable for particle simulations due to the long calculation time. For this reason the normal electric field component $E_z$, the transversal electric field component $E_x$ and the magnetic field component $B_y$ are described with the suitable approximations

$$E_x(x, z, t) = -\sqrt{2} \frac{\pi U_0 D_e}{r_D^2} \frac{\partial \rho}{\partial x_n}\left(\frac{x}{r_D}, t\right) \cdot H\left(\left|\frac{z}{r_D}\right| - \frac{\Delta a}{2r_D}\right) \cdot \exp\left[-\left(\left|\frac{z}{r_D}\right| - \frac{\Delta a}{2r_D}\right)/\sqrt{2}\right] , \quad (4)$$

$$E_z(x, z, t) = -\frac{\pi U_0 D_e}{r_D^2} \rho\left(\frac{x}{r_D}, t\right) \cdot H\left(\left|\frac{z}{r_D}\right| - \frac{\Delta a}{2r_D}\right) \cdot sgn\left(\frac{z}{r_D}\right) \cdot \exp\left[-\left(\left|\frac{z}{r_D}\right| - \frac{\Delta a}{2r_D}\right)/\sqrt{2}\right] , \quad (5)$$

$$B_y(x, z, t) = -\frac{2\pi^2\sqrt{2}}{c} \frac{U_0 D_e}{r_D^2} \rho\left(\frac{x}{r_D}, t\right) \cdot v_L(t) \cdot H\left(\left|\frac{z}{r_D}\right| - \frac{\Delta a}{2r_D}\right) \cdot sgn\left(\frac{x}{r_D}\right) \cdot sgn\left(\frac{z}{r_D}\right) \cdot \exp\left[-\left(\left|\frac{z}{r_D}\right| - \frac{\Delta a}{2r_D}\right)/\sqrt{2}\right], \quad (6)$$

where $U_0 = k_B T_e/e$, $sgn(x)$ is the signum-function and $H(z)$ the Heaviside step function [10].

The proposed field functions [Eq. (4), (5) and (6)] have been applied in Ref. [10] to explain the streak deflectometry measurements of Figs. 4(a) - 4(c). In this context Eq. (4), (5) and (6) have been multiplied with the additional factor $exp(-t/\tau_{ced})$ in order to account for the charge equilibration of the polymer foil on a time scale $\tau_{ced} \approx 100$ ps. Furthermore $v_L(t)$ was defined by $v_L(t) = (1/\omega_{pef}) \cdot dL(t)/dt$ and $D_e = D_{e0} = L_n(0) r_{D0}$, as well as $r_D = r_{D0}$ were applied in the conducted particle simulations. Using these expressions it became possible to reproduce the streak deflectometry measurement in longitudinal imaging configuration [Fig. 4(a)]. However, accurate results were only obtained by assuming relatively high values for the hot electron temperature $T_e$ in the order of 13 MeV and by setting the initial radius of the electron spot size $D_e$ to a value $> 10$ μm. This is relatively high in comparison to the radius of the laser spot size $r_L \approx 1.5$ μm (FWHM). In addition it was not possible to explain both the longitudinal and transversal measurements with one parameter set on the basis of Eq. (4), (5) and (6).

In order to achieve a more comprehensive explanation for the experimental results in both imaging configurations the proposed model [13] has been modified and developed further in this paper. This extended version will be referred to as model A in the following. The principle idea of model A is to find an approximation for the temporal evolution of the Debye length $r_D(t)$ and to include this time dependence into Eq. (4), (5) and (6).

The approach developed here connects the expansion of the electron spot size

$$D_e(t) = L_n(t) r_{D0} \quad (7)$$

with the temporal evolution of the hot electron density

$$n_e(t) = \frac{N_e}{\pi D_e(t)^2} \frac{1}{2r_D(t)}, \quad (8)$$

where $N_e$ is the number of hot electrons in a cylinder with the length $2r_D(t)$ and the radius $D_e(t)$. Using the definition of the Debye radius $r_D(t) = \sqrt{\varepsilon_0 k_B T_e / e^2 n_e(t)}$ Eq. (8) can be rearranged:

$$n_e(t) = \frac{N_e^2 e^2}{4\pi^2 \varepsilon_0 k_B T_e \, D_e(t)^4} \qquad (9)$$

and the time dependence of the Debye length can be expressed as

$$r_D(t) = r_{D0} \left(\frac{D_{et}(t)}{D_{e0}}\right)^2. \qquad (10)$$

In the framework of model A Eq. (4), (5) and (6) multiplied with the decay factor $exp(-t/\tau_{ced})$ are used for the field description. However, the time dependence of the spot size as well as temporal evolution of the Debye radius are included in these equations via $D_e = D_e(t)$ [Eq. (7)] and $r_D = r_D(t)$ [Eq. (10)], respectively. In order to make model A applicable for multi-dimensional particle simulations its field description is extended to three spatial dimensions, by assuming cylindrical symmetry. As a result the normal electric field component $E_{N(A)}(r,z,t) = E_z(r,z,t)$, the radial electric field component $E_{R(A)}(r,z,t) = E_x(r,z,t)$ and the toroidal magnet field $B_{T(A)}(r,z,t) = B_y(r,z,t)$, with $r = \sqrt{x^2 + y^2}$. The spatial dependence of $E_{N(A)}(r,z,t)$ in z-direction (at $r = 0$) and its temporal evolution is illustrated in Fig. 3(a).

## B. Model B

Model B provides an analytical description of the electric field structure of a moving ion front. It is based on a one-dimensional model [14] that describes the isothermal plasma expansion into vacuum and explains the ion acceleration as the result of charge separation. In between a positively and a preceding negatively charged region an electric field is created, which peaks at a definite position. Since the accelerated ions accumulate at this position, an ion front is generated, which propagates with increasing velocity.

The spatial dependence of the field structure is described by a constant plateau region followed by a steep rise up to the front field peak and a decay of the expanding Gaussian-shaped front. As an approximation for the spatial decay of the front region a function with the proportionality [17] $E_{\text{front}}(z) \propto (1 + z/l_i)^{-1}$ is used, where the parameter $l_i$ determines the characteristic spatial scale of the decay. The electric field amplitudes of the front peak $E_f$ and the plateau region $E_p$, respectively, depend on time and are described by the formulas [10,17]

$$E_f = \frac{2\,\mathcal{E}_c}{\sqrt{2 e_m + t^2/\tau_i^2}} \qquad (11)$$

$$E_p = \frac{2\,\mathcal{E}_c}{\sqrt{2 e_m + t^2/\tau_i^2}}, \qquad (12)$$

$$\mathcal{E}_c = \sqrt{\frac{n_{e0} k_B T_e}{\varepsilon_0}}, \qquad (13)$$

where $\tau_i = 1/\omega_{pi}$ represents the reciprocal value of the ion plasma frequency $\omega_{pi} = \sqrt{n_{e0} Z e^2/m_i \varepsilon_0}$ and determines the characteristic scale of the temporal decay. The time dependent position of the ion front, which separates the plateau and front region, is described [14] by

$$z_{\text{front}} = 2\sqrt{2e_m}\, r_{D0} \left(\left(\tau \ln\left(\tau + \sqrt{\tau^2 + 1}\right) + \sqrt{\tau^2 + 1}\right) + 1\right), \tag{14}$$

where $\tau = t/(2e_m \tau_i)$ and $e_m = exp(1)$. In this isothermal description the plasma is described with one electron temperature $T_e$ and does not cool down while expanding. For this reason the model intrinsically overestimates the velocity of the ion front, which has been shown experimentally [15,18]. In addition the front velocity $v_{\text{front}}$ diverges for $t \to \infty$, which can be derived from Eq. (14). Therefore a maximum front velocity $v_{f,\text{max}}$ is introduced [10], which limits the acceleration $v_{\text{front}} \leq v_{f,\text{max}}$ and can be determined experimentally.

Model B only provides a description for the electric field component $E_N$ normal to the target surface (z-direction), but considers neither radial electric field components nor magnetic fields. So far only a one dimensional spatial dependence of the field structure $E_{N(B)}(z,t)$ is proposed here. In order to use model B in multi-dimensional particle simulations the field description is extended to three spatial dimensions [15], by assuming a Gaussian decay of the field amplitude $E_{N(B)}(r,z,t) = E_{N(B)}(z,t) \cdot exp(-(r/l_r)^2)$ in radial direction, where $r = \sqrt{x^2 + y^2}$. The z-dependence of $E_{N(B)}(r,z,t)$ (at $r = 0$) and its temporal evolution is shown in Fig. 3(b).

### C. Model C

Model C is a combination of model A and model B and provides an analytical field description, which can reproduce the measured features of both imaging configurations. In model C the radial electric field component $E_{R(C)}(r,z,t)$ and the toroidal magnetic field are $B_{T(C)}(r,z,t)$ are evaluated in the same way as in model A. However, for the calculation of the normal field component $E_{N(C)}(r,z,t)$ either model A or model B is used depending on two selection rules. Firstly, within the plateau region ($z < z_{\text{front}}$) the description always follows model A. In other words the equation $E_{N(C)}(r,z,t) = E_{N(A)}(r,z,t)$ applies if $z < z_{\text{front}}$ [cf. Eq. (14)]. Secondly, in the front region ($z \geq z_{\text{front}}$) the description of model A is only used if the field amplitude of model A is higher than the amplitude of model B, that means if $E_{N(A)}(r,z,t) > E_{N(B)}(r,z,t)$. This dependence is shown in Fig. 3(c) illustrating the z-dependence of $E_{N(C)}(r,z,t)$ (at $r = 0$) at different moments in time. As a result the different field components of Model C can be summarized using the equations

$$E_{N(C)}(r,z,t) = \begin{cases} E_{N(A)}(r,z,t), & z < z_{\text{front}} \\ \max\left[E_{N(A)}(r,z,t), E_{N(B)}(r,z,t)\right], & z \geq z_{\text{front}} \end{cases}, \tag{15}$$

$$E_{R(C)}(r,z,t) = E_{R(A)}(r,z,t), \tag{16}$$

$$B_{T(C)}(r,z,t) = B_{T(A)}(r,z,t). \tag{17}$$

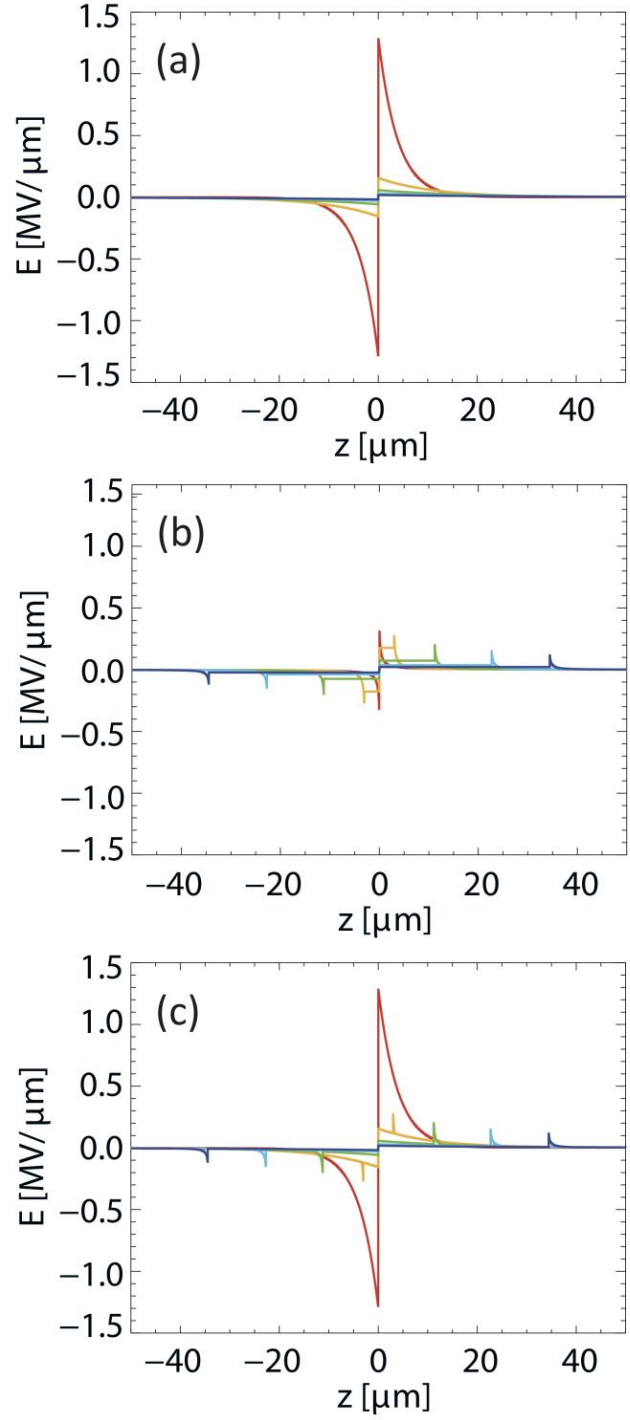

FIG. 3. Spatial distribution of the normal electric field component $E_N(r = 0, z, t)$ in z-direction at different times (red: 0 ps, orange: 0.5 ps, green: 1 ps, light blue: 1.5 ps, dark blue: 2 ps. For the field calculations the parameters of Table I and Table II have been used: (a) Model A, (b) Model B, (c) Model C

## V. SIMULATION AND DISCUSSION

In order to test the applicability range of the presented models and to draw conclusions on the laser-induced fields three-dimensional particle simulations have been conducted. For this purpose a particle tracer [19] has been used to calculate the trajectories of protons that propagate through a simulated streak deflectometry [10] including the field distribution according to one of the proposed model descriptions.

In the case of model A the determination of the parameters $D_{e0}, T_e, N_e, \sigma_{\text{cold}}, \nu_{\text{ef}}$ and $\tau_{\text{ced}}$ enables the calculation of the fields on the basis of Eq. (1) - (10). The radius of the initial electron spot size $D_{e0}$ was set to the value of the laser waist size $r_L$ (the radius where the intensity decreases by a factor of $e^{-2}$ against the maximum value). The mean laser intensity within this radius is denoted by $I_L$ and was used for the calculation of the ponderomotive potential $\phi_p$, which allows for an estimation of the hot electron temperature $k_B T_e \approx \phi_p(I_L)$ [20]. The initial number of hot electrons $N_e$ in the cylinder with the length $2r_D(t)$ and the radius $D_e(t)$ [cf. Eq. (8)] can be estimated via the formula $N_e = \eta_L \epsilon_L / (k_B T_e)$, where $\epsilon_L = I_L \pi r_L^2 \tau_L$ is the laser energy and $\eta_L$ the absorption coefficient. The conductivity of cold electrons $\sigma_{\text{cold}}$ is set to a value of approximately $\omega_{pec}/4\pi$ [13], where $\omega_{pec}$ is the cold electron frequency and depends on the cold electron density $n_{ec}$ via $\omega_{pec} = \sqrt{n_{ec} e^2 / \varepsilon_0 m_e}$. The parameters $\eta_L, \nu_{\text{ef}}$ and $\tau_{\text{ced}}$ were treated as free variables and chosen within reasonable boundaries in order to achieve a good reproduction of the experiment. Based on the applied model variables and Eq. (1) - (10) relevant field and plasma parameters were calculated. All variables and parameters are summarized in Table I. The simulation results using model A are depicted in Figs. 4(d) - 4(f). The comparison to the experiment reveals that model A allows for an accurate reproduction of the streak deflections in both imaging configurations.

Using model A with the parameters of Table I, a magnetic field amplitude of over $10^4$ T is obtained. However, the influence of the magnetic field on the proton deflection is negligible compared to the impact of the radial electric field.

Model A can also explain the measured cut-off energies of protons (between 2.4 MeV and 3.0 MeV) [10] that originate directly from the polymer foil. If the acceleration of a single proton in the field distribution of model A is calculated numerically, using the parameters of Table I, a final energy of 2.6 MeV is obtained. However model A fails to reproduce the shift of the deflection maximum to higher probing times $t_p$ with increasing distance $|x_F|$, which has been observed experimentally [cf. Figs. 3(c) and 3(f)] for larger distances ($|x_F| \gtrsim 700$ μm).

Regarding model B, the parameters $T_e, \tau_i, v_{f,\text{max}}, l_r$ and $l_s$ must be determined to calculate the normal electric field component $E_{N(B)}(r, z, t)$ within the particle simulation. The hot electron temperature $T_e$ is approximated via $k_B T_e \approx \phi_p(I_L)$ and the characteristic decay time $\tau_i$ is set to a value of approximately $1/\omega_{pi}$. The maximal expansion velocity $v_{f,\text{max}}$ is determined on the basis of the perturbation of the deflection maximum in Fig. 4(c), which propagates in space and time and has a slope that represents the expansion velocity. The remaining parameters $l_r$ and $l_s$ have been varied in order to achieve a good agreement with proton streak deflectometry measurements of Figs. 4(a) - 4(c). All variables and model parameters are listed in Table II. As the simulation results show, model B is not suited to explain the measured streak deflections in longitudinal imaging configuration [cf. Fig. 4(g)]. Also in transversal configuration [Figs. 4(h) and 4(i)] the agreement between experiment and simulation is not very accurate. Nonetheless model B can qualitatively explain the temporal evolution of the observed streak perturbations in Fig. 4(i) at larger distances ($|x_F| \gtrsim 700$ μm), which affirms the assumption that these perturbations are caused by the field of a moving proton front.

Figs. 4(a) - 4(c) show the calculated streak deflections when the field description of model C is applied within the simulation. Apparently the proposed combination of the two models [Eq. (15) - (17)] enables an extensive and accurate reproduction of the experimental results [Figs. 4(a) - 4(c)]. Using model C, not only single and selected measurements can be reproduced by the particle simulation, but also (and at the same time) measurements with varied experimental parameters and measurements that were obtained using different imaging configurations.

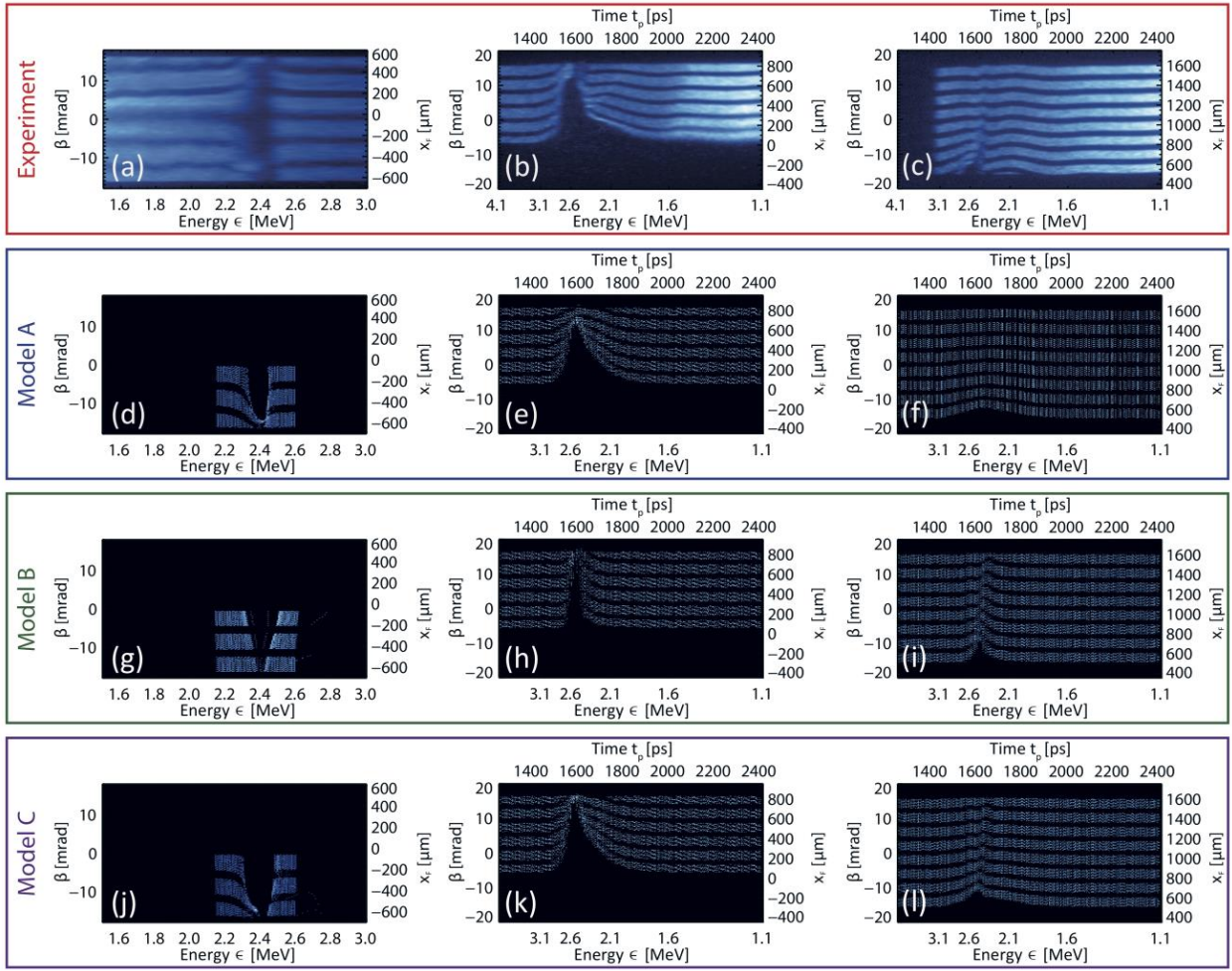

FIG. 4. Experimental [(a)-(c)] and simulated [(d)-(l)] streak deflections in longitudinal [(a),(d),(g),(j)] and transversal [(b),(c),(e),(f), (h),(i),(k),(l)] imaging configuration.

TABLE I. Variables and simulation parameters of model A and model C.

| Parameter | Value | Unit |
|---|---|---|
| $D_{e0}$ | $4 \cdot 10^{-6}$ | m |
| $T_e$ | $1.22 \cdot 10^{10}$ | K |
| $N_e$ | $2.15 \cdot 10^9$ | 1 |
| $\sigma_{cold}$ | $7.36 \cdot 10^{14}$ | 1/s |
| $\nu_{ef}$ | $1 \cdot 10^{-4}$ | 1 |
| $\tau_{ced}$ | $7.5 \cdot 10^{-11}$ | s |
| $\phi_p$ | $1.68 \cdot 10^{-13}$ | J |
| $\phi_{p,\text{MeV}}$ | 1.05 | MeV |
| $r_{D0}$ | $2.72 \cdot 10^{-6}$ | m |
| $\eta_L$ | $8 \cdot 10^{-4}$ | 1 |
| $I_L$ | $1.8 \cdot 10^{19}$ | W/cm$^2$ |
| $\epsilon_L$ | 0.452 | J |
| $\tau_L$ | $5 \cdot 10^{-14}$ | s |
| $r_L$ | $4 \cdot 10^{-6}$ | m |
| $\omega_{pec}$ | $1.26 \cdot 10^{16}$ | 1/s |
| $n_{ec}$ | $5 \cdot 10^{28}$ | 1/m$^3$ |
| $L_{n0}$ | 1.47 | 1 |
| $\kappa$ | $5.13 \cdot 10^{-2}$ | 1 |
| $\omega_{pef}$ | $1.58 \cdot 10^{14}$ | 1/s |
| $n_{e0}$ | $7.86 \cdot 10^{24}$ | 1/m$^3$ |
| $\Delta a$ | $3 \cdot 10^{-8}$ | m |
| $\eta_{e0}$ | 9.24 | 1 |
| $S_L$ | 0.3 | 1 |
| $U_0$ | $1.05 \cdot 10^6$ | V |

TABLE II. Variables and simulation parameters of model B and model C.

| Parameter | Value | Unit |
|---|---|---|
| $T_e$ | $1.22 \cdot 10^{10}$ | K |
| $\tau_i$ | $350 \cdot 10^{-15}$ | s |
| $\omega_i$ | $3.69 \cdot 10^{12}$ | 1/s |
| $v_{f,\text{max}}$ | $2.35 \cdot 10^7$ | m/s |
| $l_r$ | $400 \cdot 10^{-6}$ | m |
| $l_s$ | $0.25 \cdot 10^{-6}$ | m |

## VII. PICOSECOND EXPERIMENTS

Figs. 5(a) and 5(b) show two processed streak deflectometry measurements in longitudinal imaging configuration. Both measurements were performed using identical experimental parameters (cf. experimental setup). However in case of Fig. 5(a) the position of the grating compressor, which provided the shortest pulse length, was changed by a distance of 10 mm. As a result, a pulse length $\tau_L$ of approximately 2.7 ps was generated, leading to an averaged intensity level within the beam waist of $3.3 \cdot 10^{17}$ W/cm² [$5.8 \cdot 10^{17}$ W/cm² (FWHM)]. In addition the variation of the grating position introduced a time delay between the pump and probe laser pulse, which resulted in a change of the associated t-pump energy $\epsilon_{pump}$ from 2.4 MeV to 1.227 MeV.

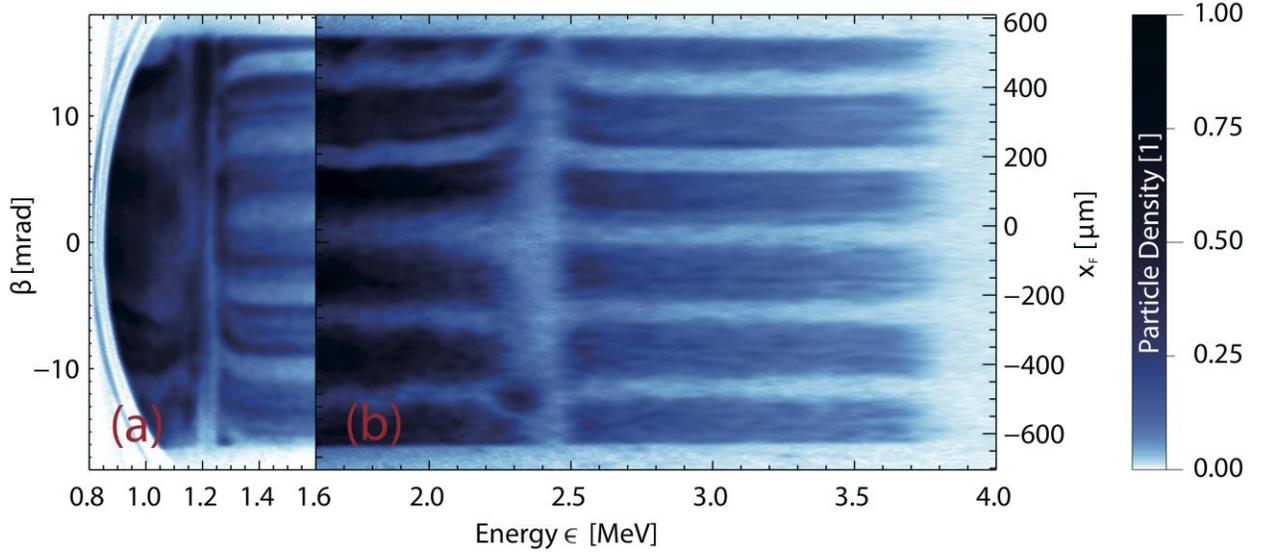

FIG. 5. Comparison between long-pulse-induced (a) and short-pulse-induced (b) streak deflections.

Both in the femtosecond case [Fig. 5(b)] and in the picosecond case [Fig. 5(a)] prominent density gaps appear in the recorded proton streaks within the energy range around the value of the t-pump energy $\epsilon_{pump}$. This can be explained by the decelerating and accelerating effect of an ambipolar electric field around the polymer foil, which leads to a redistribution of the proton energies around $\epsilon = \epsilon_{pump}$ [12]. Thus, a clear allocation between $\epsilon$ and $t_p$ is not possible. Nevertheless, in principle higher proton energies correspond to lower probing times and thus reflect an earlier stage of the field evolution at the polymer foil.

In the picosecond case, the proton streaks at the low energy side of the gaps appear broadened, deformed and blurred, making the interpretation difficult. However, at the high-energy side a clear bending of the proton streaks becomes visible in form of a focusing effect. Here the streak deflections are directed inwards, that is towards the center of the laser plasma interaction at $|x_F| = 0$.

To see this, one should keep in mind that the mean initial emission angle of a single trace can be deduced by its position at energies $\epsilon \gg \epsilon_{pump}$ (cf. Section III). In case of Fig. 5(a) the energy $\epsilon = 1.6$ MeV allows for a relatively accurate estimation. At lower energies ($\epsilon \approx 1.25$ MeV) the absolute values of the mean propagation angles (of the streaks) are clearly reduced in comparison to their initial emission angles $\beta$.

In contrast, the proton streak deflections in the femtosecond case are exactly opposite, pointing outwards. The induced and controllable change of the deflection direction of protons within a certain energy interval is a significant experimental finding. Regarding the picosecond case, the application of a Thomson slit spectrometer allowed for a clear determination of the magnetic field configuration at the rear side of the polymer foil. Depending on its kinetic energy, each probing proton has a position either before (z < a) or behind (z > a) the polymer foil [cf. Figure 1 and 6], when the deflecting fields are initiated at the time $t_{pump}$. Due to the definition of the t-pump energy $\epsilon_{pump}$ (cf. Section Experimental Results) all protons with energies $\epsilon > \epsilon_{pump}$ have already passed the polymer foil and are located at a position behind the rear side of the foil (z > a) at the time $t_p = t_{pump}$.

In the picosecond case, the inward bending of the proton streaks at the high energy side of the gap just appears within the energy range $\epsilon > \epsilon_{pump}$. Therefore, these protons reflect the exclusive influence of the extended fields and the associated force acting on the rear side of the foil. Conversely, this means that none of these protons are influenced by the electric and magnetic field distribution acting at the front side of the polymer target.

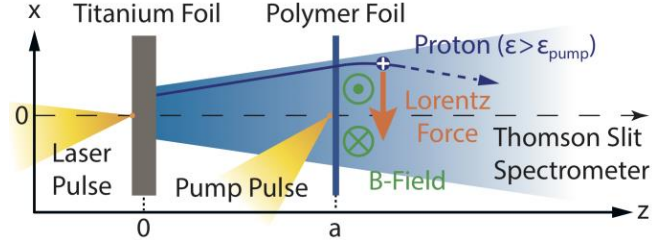

FIG. 6. Deflection of a single proton at the rear side of the polymer foil due to the influence of the $(\vec{v} \times \vec{B})$-term of the Lorentz force

From PIC-simulation we know that fields located at rear rand front-side of the target do not extend to the opposite side. This situation is illustrated in Fig. (6) showing the direction of the acting Lorentz force on a proton behind the rear side of the foil (z > a) at the time $t_p = t_{pump}$. Due to the charge up of a thin foil during a laser plasma interaction the radial electric field component must point outward. This is just opposite to the marked direction of the net force that leads to the observed focusing effect. Therefore, the radial electric field component allows no explanation for the measured inward deflection.

In principle also the longitudinal electric field could be responsible for a change of the propagation direction. After the acceleration process at the titanium foil each proton has a certain initial ejection angle $\beta = \tan(v_x/v_z)$ depending on its longitudinal and transversal velocity components. If the proton is reaccelerated by the longitudinal electric field of the polymer foil its longitudinal velocity component can be increased while its radial velocity component remains constant. This leads to a reduction of the ratio $|v_x/v_z|$ and thus of the absolute value of the measured angle $\beta$. In principle this would lead to an inward bending of the proton traces.

However, based on the measurement in Fig. 5(a) one can estimate that this mechanism is not sufficient to explain the observed proton defection quantitatively. If one regards, for instance, the lower edge of second proton trace from above one can deduce that $\beta$ changes from its initial value of ~9 mrad to a value of ~7 mrad. In addition an effective energy gain $\Delta \epsilon$ of approximately 0.05 MeV can be deduced from the width of the energy gap.

Given a proton with the initial energy $\epsilon_0 = \epsilon_{pump} = 1.227$ MeV and the initial ejection angle $\beta_0 = 9$ mrad, one can calculate the initial longitudinal velocity $v_{z0} = 1.532 \cdot 10^7$ m/s and transversal velocity component $v_{x0} = 1.379 \cdot 10^5$ m/s. After the interaction the proton has the energy $\epsilon_1 = \epsilon_0 + \Delta \epsilon \approx 1.277$ MeV and the longitudinal velocity component $v_{z1}$ has a value of ~1.563 $\cdot 10^7$ m/s. Assuming that the transversal velocity component $v_{x1}$ remains unchanged during the interaction ($v_{x1} = v_{x0}$), one can estimate that the propagation angle $\beta_1$ after the interaction has an approximate value of 8.824 mrad. The change of the emission angle can be calculated by $\Delta \beta = \beta_0 - \beta_1$ and has the value ~0.176 mrad, which cannot explain the observed change of ~2 mrad.

Therefore, the $(\vec{v} \times \vec{B})$-term of the Lorentz force must be responsible for the observed phenomenon and the direction of the acting force in Fig. (6). Hence, the term must be dominant against the radial electric field component, which means that $|\vec{v} \times \vec{B}| > |\vec{E}_R|$. In addition, the polarity of the laser induced toroidal magnetic field must be consistent with the marked direction in Fig. (6), otherwise a defocusing effect would occur. This phenomenon is observed for the first time in connection with high intensity laser pulses ($I_L > 10^{17}$ W/cm$^2$) with high temporal contrast and solid ultra-thin foil targets. In recent investigations [21,22] of the magnetic field structure generated on thin foils using high intensity lasers ($I_L \sim 10^{19}$ W/cm$^2$) just the opposite magnetic polarity was found. However, this polarity [i.e. opposite to the marked direction in Fig. (6)] is also in agreement with the analytical description of model A, explaining the longitudinal streak deflection measurements in connection with femtosecond laser pulses.

In the case of laser irradiated thin foils, the polarity of the thermoelectric magnetic field generation [23,24,25,26] is just the opposite to the observed polarity in the picosecond case as illustrated in Fig. (6). This indicates that another mechanism is responsible for the magnetic field generation. In principle the observed magnetic field polarity is possible due to DC currents in steep density gradients driven by temporal variations of the ponderomotive force [27,28,29]. Another explanation is the occurrence of hot electron currents, which are either directed into the target [30,31] or along the target surface [32,33,34]. So far magnetic fields with a polarity opposite to the orientation of a thermoelectric field were only observed experimentally in the interaction of relativistic intense laser pulses that propagate in a pre-ionized plasma [35]. This raises the question if an under-critical plasma could be generated due to the picosecond pulse interaction.

The observed proton deflection pattern of Fig. 5(a) at the high-energy side of the gap can be quantitatively reproduced with a numerical particle simulation [10]. The simulation result confirms the measured polarity of the magnetic field at the rear side of the target.

However, a clear determination of the magnetic field generation mechanism is not possible. Therefore, further theoretical research is required to find the physical mechanism that leads to the creation of the observed magnetic field.

## IX. CONCLUSION

In this paper we studied the dynamics and structure of the electric and magnetic fields that were generated on ultra-thin polymer foils ($30 - 50$ nm) when irradiated with high intensity femtosecond ($10^{19} - 10^{20}$ W/cm$^2$) and picosecond ($\sim 10^{17}$ W/cm$^2$) laser pulses using ultra-high laser contrast ($10^{10} - 10^{11}$). For this purpose the proton streak deflectometry method has been applied and developed further. We could show that the use of ultra-high contrast laser pulses significantly increased the reproducibility of experiments and improved the accuracy of the applied imaging method in comparison to former experiments. Further advancement was realized by applying the method in two different imaging configurations. It enabled us to gather complementary information about the investigated field structure. In particular the influence of different field components (parallel and normal to the target surface) and the impact of a moving ion front could be studied in detail. Regarding the experiments with femtosecond laser pulses these particular measured features could be explained using two different models providing an analytical description of the laser induced fields. Their ability to reproduce the streak deflectometry measurements was tested on the basis of three-dimensional particle simulations. Simulating the experiment in two different imaging configurations expanded our knowledge of the application range and limitations of the single models and enabled the further development of the investigated analytical model descriptions itself. A modification and combination of the two models allowed for an extensive and accurate reproduction of the experimental results in both imaging configurations. This way, not only single and selected measurements could be reproduced by the particle simulation, but also (and at the same time) measurements with varied experimental parameters and measurements that were obtained using different imaging configurations. In this regard, the described methodological approach might offer a new path for a comprehensive reconstruction of the spatial and temporal field distribution of laser-plasma interactions.

In addition, we could demonstrate that a controlled change of the pulse duration can be used to manipulate the propagation direction of the proton beam within a certain energy range. The change from 50 femtosecond to 2.7 picosecond laser pulses (with the same energy content) led to a transition of the dominating force acting on the probing proton beam at the rear side of the polymer foil. In the picosecond case the ($\vec{v} \times \vec{B}$)-term of the Lorentz force dominated over the action of the $\vec{E}$-field and was responsible for the direction of the acting force. The applied imaging method allowed for an unambiguous determination of the magnetic field polarity at the rear side of the ultra-thin foil, revealing an unexpected orientation of the field that is just opposite to the polarity of the thermoelectric magnetic field generation.

**ACKNOWLEGMENTS**

The research leading to these results has received funding from Deutsche Forschungsgemeinschaft within the program CRC/Transregio 18 and from LASERLABEUROPE (Grant Agreement No. 284464, EC's Seventh Framework Program).A.A.A. acknowledges the provided computation resources of JSC at project HBU15. PVN acknowledges the support by the Institute for Basic Science under IBS-R012-D1.